\documentstyle[12pt]{article}
\def\be{\begin{equation}}
\def\ee{\end{equation}}

\author{Hans - J\"urgen Schmidt}

\title{Non-trivial Solutions of the Bach Equation Exist}

\date{}
\begin{document}
\maketitle

\centerline{Universit\"at Potsdam, Institut f\"ur Mathematik, Am
Neuen Palais 10} 
 \centerline{D-14469~Potsdam, Germany,  E-mail:
 hjschmi@rz.uni-potsdam.de}

\bigskip

\begin{abstract} 
We show that solutions of the Bach equation exist
 which are not conformal Einstein spaces. 
\end{abstract}

In connection with fourth order gravitational field
 equations, cf.  e.g. [1, 2] where the breaking of conformal invariance
 was discussed, the original {\sc Bach} equation,
\be  
B_{ij} = 0 \, ,              %(1
\ee
enjoys current interest. Eq. (1) stems from a Lagrangian 
$$
L = \frac{1}{2} \sqrt{-g} C_{ijkl} C^{ijkl}   \,  ,
$$
and variation gives, cf. {\sc Bach} [3],
$$
 \frac{1}{\sqrt{-g}} \delta L / \delta g^{ij}     
  = B_{ij} = 2 C^{a \ \  b}_{\ ij \ ;ba}
 +          C^{a \ \  b}_{\ ij } R_{ba}
       \,  .
$$
An Einstein space,
\be
     R_{ij} = \lambda g_{ij} \,  , %(2)
\ee
is always a solution of the 
{\sc Bach}
 equation (1). But eq. (1) is conformally invariant, and therefore, 
each metric, which is conformally related to an Einstein space, 
fulfils eq. (1), too. We call such solutions  trivial ones.

Now the question arises whether non-trivial solutions of the {\sc Bach}
 equation (1) do or do not exist, and the
 present note will give an affirmative answer. 
As a by-product, some conditions  will be
 given under which only trivial solutions exist. 
Observe that eq. (1) is conformally invariant whereas eq. (2) is not. 
Therefore, a simple counting of degrees of freedom does not suffice.

Because the full set of solutions of eq. (1)
 is not easy to describe, let us consider some homogeneous cosmological
models.
 Of course, we have to consider anisotropic ones, because all 
Robertson-Walker models are trivial solutions of eq. (4). 
Here, we concentrate on the diagonal Bianchi type I models
\be
ds^2 = dt^2  - a^2_i  dx^{i^2}    %(3>
\ee
with Hubble parameters $h_i = a_i^{-1} da_i/dt$, $h=\Sigma h_i$
and anisotropy parameters $m_i = h_i 
-    h/3$. The Einstein spaces of this kind are described in [4], 
for $\lambda =  0$ it is just the Kasner
metric $ a_i =   t^{p_i}$,
 $\Sigma p_i = \Sigma p_i^2 =   1$. 
All these solutions have the property that the quotient of two
anisotropy parameters,
 $m_i/m_j$, (which 
equals $ (3p_i - 1)/(3p_j - 1) $  for the Kasner metric)
 is independent
of $t$, and this property is a conformally invariant one. 
Furthermore, it holds: {\it  A solution of eqs. (1),
(3) is a trivial one, if and only if the quotients   $m_i/m_j$    are
constants.}

Restricting now to axially symmetric Bianchi type I models, 
i.e., metric  (3) with $ h_1 = h_2$, the
identity $\Sigma m_i = 0$
 implies $m_1/m_2 = 1$, $m_3/m_1 =   m_3/m_2 =   -2$ ,
 i.e., {\it each axially symmetric Bianchi
type I  solution of eq. (1) is  conformally related to an Einstein space.} 
 (Analogously, all static 
spherically symmetric  solutions of eq. (1)  are trivial ones, cf. [5].)

Finally, the existence of a solution of eqs. (1), (3) with a non-constant
 $ m_1/m_2 $ will be shown. For
the sake of simplicity we use the gauge condition $h = 0$,
 which  is possible because of  the
conformal invariance of eq. (1). 
Then the 00 component and the 11 component of eq. (1) are
sufficient to determine the unknown functions $h_1$  and $h_2$;
 $h_3 = -h_1 - h_2$
 follows from the gauge condition. 
Defining 
$r = (h^2_1 + h_1h_2 + h_2^2)^{1/2}$  and $p = h_1/r$, 
 eq. (1) is equivalent to the system                
\be
3d^2(pr)/dt^2 =  8pr^3 + 4c, \quad c = {\rm const.,} \quad   p^2 \le  4/3
\,  ,  %(4)
\ee
\be 
9(dp/dt)^2 r^4 =   [2r d^2 r/dt^2 - (dr/dt)^2 - 4r^4] (4r^2 - 3p^2r^2) \,
. % (5)
\ee
As one can see, solutions with a non-constant $p$
 exist, i.e., $m_1/m_2$  is not constant for this case.

Result. {\it Each solution of the
 system (3), (4), (5) with $dp/dt \ne  0$  represents a non-trivial solution
of the {\sc Bach} equation (1).}

\medskip

The author thanks Dr. G. {\sc Dautcourt} 
 and Prof. Dr. H. {\sc Treder}  for stimulating discussions.

\medskip

\noindent 
{\bf   References}

%\medskip
\noindent 
[1] {\sc Weyl}, H.: Raum, Zeit, Materie, 4. Aufl. Berlin: Springer-Verlag
1921.

\noindent 
[2] {\sc von  Borzeszkowski}, H.; {\sc 
Treder}, H.-J.; {\sc Yourgrau}, W.: Ann. Phys. Leipzig {\bf 35}
(1978) 471.

\noindent 
[3] {\sc Bach}, R.: Math. Zeitschr. {\bf 9} (1921) 110.

\noindent 
[4] {\sc Kramer},
 D., et al.: Exact solutions of Einstein's field equations. Berlin: Verlag
der 
Wissenschaften 1980, eq. (11.52).   

\noindent 
[5] {\sc Fiedler}, B.; {\sc Schimming}, R.: Rep. math. phys. {\bf 17}
(1980) 15.

\bigskip

\noindent 
Bei der Redaktion eingegangen am 16. November 1984.

\medskip

\noindent 
{\small {In this reprint we only added the abstract and 
removed  obvious misprints of the original, which
was published in 
Annalen der Physik (J. A. Barth, Leipzig). 7. Folge, Band 41, Heft 6, 1984, 
S. 435-436; 
 author's address that time:  
Zentralinstitut f\"ur  Astrophysik der AdW der DDR, 
1502 Potsdam--Babelsberg, R.-Luxemburg-Str. 17a. }}

\end{document}